\documentclass[12pt]{article}
\usepackage{epsf}
\usepackage{epsfig}
\usepackage{hyperref}
\usepackage{amssymb,amsmath,dsfont,theorem,multirow}

\makeatletter
\renewcommand\section{\@startsection {section}{1}{\z@}%
                                   {-3.5ex \@plus -1ex \@minus -.2ex}
                                   {2.3ex \@plus.2ex}%
                                   {\normalfont\large\bfseries}}

\renewcommand\subsection{\@startsection{subsection}{2}{\z@}%
                                     {-3.25ex\@plus -1ex \@minus -.2ex}%
                                     {1.5ex \@plus .2ex}%
                                     {\normalfont\bfseries}}

\makeatother

\long\def\symbolfootnote[#1]#2{\begingroup%
\def\thefootnote{\fnsymbol{footnote}}\footnote[#1]{#2}\endgroup}

\newcommand{\bea}{\begin{eqnarray}}
\newcommand{\eea}{\end{eqnarray}}
\newcommand{\be}{\begin{equation}}
\newcommand{\ee}{\end{equation}}

\newcommand{\bem}{\begin{pmatrix}}
\newcommand{\eem}{\end{pmatrix}}

\def\a{\alpha}
\def\b{\beta}

\def\d{\delta}

\def\e{\epsilon}   
\def\f{\phi}               
  
\def\g{\gamma}

\def\im{\mathrm{Im}}
\def\inf{\infty}

\def\l{\lambda}

\def\n{\nu}

\def\p{\pi}   
\def\pa{\partial}       
\def\re{\mathrm{Re}}                
\def\r{\rho}                                     
\def\s{\sigma}                                   
\def\t{\tau}
\def\th{\theta}
\def\til{\tilde}

\def\z{\zeta}

\def\G{\Gamma}

\def\L{\Lambda}
\def\O{\Omega}
\def\P{\Pi}
\def\Q{\Theta}
\def\S{\varSigma}

\def\cn{{\cal N}}

\def\cz{{\cal Z}}

\def \Z {{\mathbb Z}}
\def \C {{\mathbb C}}
\def \R {{\mathbb R}}

\def\SS{\scriptscriptstyle}
\def\bt{\bar\tau}
\def\bl{\bar\lambda}

\begin{document}
\begin{titlepage}

\begin{center}

\hfill ITFA-2009-02
\vskip 2 cm {\Large \bf  A Geometric Derivation of the\\[3mm] Dyon Wall-Crossing Group}
\vskip 1.25 cm { 
Miranda C. N. Cheng
\symbolfootnote[1]{mcheng@physics.harvard.edu}}\\
{\it Jefferson Physical Laboratory, Harvard University, Cambridge, MA 02128, USA}\\ 
{\vskip 0.3cm
and}
{\vskip 0.3cm
Lotte Hollands\symbolfootnote[2]{L.Hollands@uva.nl}\\
  {\it  Institute for Theoretical Physics, University of Amsterdam,}\\ 
 {\it Valckenierstraat 65, 1018 XE, Amsterdam, the Netherlands}
}

\end{center}

\vskip 2.5 cm

\begin{abstract}
\baselineskip=18pt  

Recently,  using supergravity analysis, a hyperbolic reflection group was found to underlie the structure of wall-crossing, or the discontinuous moduli dependence of the supersymmetric index due to the presence of walls of marginal stability, of the BPS dyons in the ${\cal N}=4$, $d=4$ compactification. In this paper we work in the regime where four-dimensional gravity decouples and we show how the presence of such a group structure can be easily understood as a consequence of the supersymmetry of a system of $(p,q)$ five-brane network, or equivalently the  holomorphicity of the Riemann surface wrapped by the appropriate M5 branes in the Euclidean M-theory frame. 

\end{abstract}
\end{titlepage}

\pagestyle{plain}
\baselineskip=19pt

\tableofcontents

\section{Introduction}
\label{Introduction}

Recent years have seen a lot of progress in the understanding of the BPS spectrum of ${\cal N}=4$, $d=4$ string theory \cite{Lopes Cardoso:2004xf}-\cite{Banerjee:2008ky}.  First and foremost, complete microscopic partition functions have  been proposed for various ${\cal N}=4$, $d=4$ string compactifications, including the so-called CHL models \cite{Dijkgraaf:1996it},\cite{Jatkar:2005bh}. These proposals for the generating function for the BPS indices of the
theories have in the meanwhile been fairly well-understood and passed all consistency checks performed so far \cite{Dijkgraaf:1996it,Lopes Cardoso:2004xf,Sen:2007vb,Dabholkar:2007vk,Cheng:2007ch}.  

Most remarkably, it has been observed that these partition functions encode the BPS indices at all points in the moduli space
 \cite{Sen:2007vb,Cheng:2007ch}. Recall that, due to the presence of walls of marginal stability in the moduli space where states could (dis)appear, the graded degeneracies of the BPS states are only piecewise constant when one changes the value of moduli at spatial infinity and in particular typically jump when a wall of marginal stability is crossed. See also \cite{Denef:2007vg,4d3d,KS} for recent related discussions in $\cn=2$ and more general context. 
 But the ${\cal N}=4$ partition functions magically know about the different degeneracies in different parts of the moduli space, provided that the relevant automorphic forms are expanded using the appropriate moduli-dependent expansion parameters. 

The surprise did not stop there. More recently, relying on supergravity analysis, it was established in \cite{Cheng:2008fc} that there is a hyperbolic reflection group $W$ underlying the phenomenon of dyon wall-crossing of the $K3\times T^2$ compactification of type II string theory in an extremely simple way.
 It was also observed that the proposed dyon degeneracy formula can be seen as associating a Verma module of a generalized Kac-Moody algebra to a given total charges and moduli. In this interpretation, the difference in BPS index arises because the highest weights of the relevant Verma modules are related to each other by an element of the Weyl group of the algebra, which coincides with the group of wall-crossing $W$ obtained from the supergravity analysis. An analogue structure is also present in a $\Z_n$-orbifolded version of the theory when $n<4$ \cite{Cheng:2008kt}.

These unexpected properties of the BPS degeneracies certainly hint at deeper structures of the theories yet to be fully uncovered. Specifically, while the properties pertaining to the intricate moduli dependence of the BPS index mentioned above have been observed within the framework of ${\cal N}=4$, $d=4$ supergravity, a microscopic understanding of these properties is clearly desirable. In particular, we would like to understand why the different indices at different points in the moduli space can be extracted from the same generating function. 
More explicitly, from the fact that  the group of wall-crossing is a subgroup of the ($\Z_2$-parity-extended) S-duality group, when the moduli cross a wall of marginal stability, the change of the BPS index can be summarized by a change of the ``effective charges" by a Weyl reflection \cite{Cheng:2008fc}. We would like to understand why the index should change in such a simple way. 

To answer the above questions we will adopt a strategy similar to the one used in a recent paper \cite{Banerjee:2008yu}. It has been long known that the dyon partition function is an object naturally associated to a genus two Riemann surface \cite{Dijkgraaf:1996it}. In particular, the Igusa cusp form $\Phi_{10}(\O)$ appearing in the dyon partition function arises naturally as the partition function of 24 chiral bosons on a genus two surface \cite{Moore:1986rh}. 
Such a genus two surface occurs in the dyon counting problem in the following way \cite{Gaiotto:2005hc,Dabholkar:2006bj}. Consider type IIB string theory compactified on $K3 \times T^2$, using the appropriate U-duality frame, the $1/4$-BPS dyons of the theory can be represented by a network of $(p,q)$-string and 5-brane bound states. Euclideanizing and compactifying the time direction in order to calculate a partition function, we obtain a system which is equivalent to Euclidean M-theory compactified on $K3\times T^2 \times T^2$, with the BPS dyon represented now as an M5 brane wrapping K3 times a genus two Riemann surface holomorphically embedded in $T^4$. Now, working in a decompactification limit in which the $K3$ manifold has large volume in string unit, the authors of \cite{Banerjee:2008yu}  have succeeded in obtaining an explicit expression for the periods of the genus two Riemann surface, which was anticipated from the earlier proposal for the moduli-dependent expansion parameters of the partition function \cite{Cheng:2007ch}. In particular, on general grounds and from earlier results we expect the Riemann surface to degenerate in a certain way when the moduli cross a wall of marginal stability \cite{Sen:2007vb,Cheng:2007ch}. 

Carrying this analysis one step further, we study the change of the surface when it goes through such a degeneration, and find that it is equivalent to a particular change of the homological cycles of the surface. 
Using the relation between the homology class in the spacetime $T^4$ of the Riemann surface wrapped by the M5 brane and the conserved charges, we see how the change of the BPS index when crossing the wall of marginal stability under consideration amounts to a change of the ``effective charges" by acting by a certain element of the hyperbolic reflection group $W$. Following such a strategy and using essentially only the supersymmetry condition, we derive the specific group structure underlying the wall-crossing of the theory, and the fact that the BPS degeneracies at different moduli are given by the same partition function. 
In particular, we see how the moduli space and its partitioning by the walls of marginal stability can be identified with the dual graph of the type IIB $(p,q)$ 5-brane network compactified on the spacetime torus, with the symmetry group of the network identified with the symmetry group of the fundamental domain of the group of wall-crossing. We hope that this microscopic derivation of the Weyl group will be a first step towards an understanding of the microscopic origin of the Borcherds-Kac-Moody algebra in the dyon spectrum.

The rest of the paper is organized as follows. In section \ref{The Five-brane Network} we review and extend the results in \cite{Sen:1997xi} and discuss in details how the $1/4$-BPS dyons are realized as a periodic network of effective strings in type IIB frame at arbitrary moduli. In section \ref{The Riemann Surface} we review and extend the results in \cite{Banerjee:2008yu} by going to Euclidean M-theory and analyze the Riemann surface wrapped by the M5 brane which makes up the $1/4$-BPS dyon. In particular we analyze the complex structure of the surface and its relation to the stability of the dyon states. Section \ref{Deriving the Group of Discrete Attractor Flow} contains most of the results in the present paper. In section \ref{The First Degeneration} we focus on one specific degeneration of the surface and analyze the change of the homology cycles under such degeneration by using a hyperelliptic model of the genus two surface. In this way we derive one of the elements of the reflection group $W$. In section \ref{The Symmetry} we study the symmetry of the hyperelliptic surface, or equivalently the symmetry of the periodic network of effective strings in the type IIB frame. In this way we obtain the other generators of the group $W$. Using these results, in section \ref{Moduli Space as the Dual Graph} we discuss how the moduli space and its partitioning by the walls of marginal stability, or equivalently the walls of degenerations of the Riemann surface, can be understood simply as being the dual graph of the periodic effective string network. We also discuss the implication of these results for the counting of BPS dyonic states, and in particular why the index simply changes by an appropriate change of the ``effective charges" when the moduli cross a wall of marginal stability. In section \ref{Conclusion} we conclude by a discussion, in particular we discuss what we cannot derive by such a simple analysis and sketch an analogous treatment for the case of the CHL models.

\section{The Five-brane Network}
\label{The Five-brane Network}
\setcounter{equation}{0}

Following Banerjee, Sen and Srivastava \cite{Banerjee:2008yu}, in this section we consider $1/4$-BPS dyons made up from a type IIB $T^2$-compactified network of effective strings which are bound states of $(p,q)$ strings and  $K3$-wrapped five-branes. Working in the limit of large $K3$ and thus heavy five-branes and using the supersymmetry condition of the network \cite{Sen:1997xi}, we will write down explicit expressions for the shape and size of the network with a given range of values of the axion-dilaton and the torus complex moduli. After that we briefly discuss how the network is realized at generic values of moduli, while leaving the details to section \ref{Moduli Space as the Dual Graph}.

First consider type IIB string theory compactified on the product of a $K3$ manifold and a 
torus which we shall call $T^2_{\SS (IIB)}
$, and two effective strings wrapping the two homological cycles of the torus. Each 
effective string is a bound state of F1 and D1 string together with NS5 
and D5 branes wrapped on K3. 

To be more specific, let's consider the following charges. Suppose 
we have the $Q$ effective string, which is a bound string of a $
(n_1,n_2)$ string together with a $K3$-wrapped $(q_1,q_2)$ five-brane, wrapping the $A$-cycle of the $T^2_{\SS (IIB)}$. Wrapping the 
$B$-cycle is what we call the $P$  effective string, which is a bound 
string 
of a $(m_1,m_2)$ string together with a $K3$-wrapped $(p_1,p_2)$ five-brane. The three T-duality invariants corresponding to this charge 
configuration are given by
\be \label{charges_initial}
Q^2 = 2\sum_{i=1}^2 n_i q_i\;\;,\;\;P^2 =  2\sum_{i=1}^2 m_i p_i\;\;,\;\;Q
\cdot P = \sum_{i=1}^2 ( m_i q_i + n_i p_i)\;.
\ee

In the limit of large $K3$, the tension of the $Q$- and $P$- string are given by
\be\label{central_charge_vector}
T_Q =  \, q_1-\bar\l q_2 \quad,\quad T_P = p_1 - \bl p_2
\ee
rescaled by a factor of the volume of $K3$ in ten-dimensional Planck unit $V_{K3}^{\SS (P)}= V_{K3}\l_2$.
Here $V_{K3}$ denotes the the volume of $K3$ in string unit, and $-\bar\l = -\l_1 + i\l_2 $ is the axion-dilaton of the type IIB theory. In particular, the string coupling is given by $g_s = \l_2^{-1}$.  Similarly, we will denote by $-\bt = -\t_1 + i \t_2 $ and $R_B^2 \t_2 $ the complex structure 
and the area of the type IIB torus $T^2_{\SS (IIB)}$ respectively. 

Using the above convention,  the $SL(2,\Z)\times SL(2,\Z)$ symmetry of the 
theory acts as
\be
\t \to \frac{a\t+b}{c\t+d} \;,\;\; \bem Q \\ P \eem \to \bem a &b \\ c & 
d\eem \bem Q\\ P \eem\quad,\;\; \g =\bem a &b \\ c & d\eem \in SL(2,\Z)
\;,
\ee
and independently
\be\label{IIB_S_dual}
\l \to \frac{a'\l+b'}{c'\l+d'} \;,\;\; \bem \G_1 \\ \G_2 \eem \to \bem a' &b' \\ c'& d'\eem \bem \G_1\\ \G_2 \eem\quad,\;\; \g'=\bem a' &b' \\ c' & d'\eem\in 
SL(2,\Z)
\ee
for all $(\G_1,\G_2)$ strings or five-branes.
The second symmetry is the type IIB S-duality, while the first symmetry 
is the modular transformation of the type IIB torus $T^2_{\SS(IIB)}$, 
which is mapped to the S-duality of the heterotic string under string duality.

It will turn out to be useful to organize the above complex structure of the  torus $T^2_{\SS(IIB)}$ and the type IIB axion-dilaton field in terms of 
the following $2\times 2$ symmetric real matrices
\be
{\cal M}_\t = \frac{1}{\t_2} \bem|\t|^2 & \t_1 \\ \t_1 & 1\eem\quad,\quad
{\cal M}_\l = \frac{1}{\l_2} \bem|\l|^2 & \l_1 \\ \l_1 & 1\eem\;,
\ee
which transforms as ${\cal M}_\t \to \g{\cal M}_\t \g^T$ and ${\cal M}_\l \to \g'{\cal M}_\l \g'^T$ under the above $SL(2,\Z)\times SL(2,\Z)$ transformation. Furthermore, we will use the following standard metric on the space of $2\times 2$ 
symmetric real matrices $X$ 
\be\label{Lorenztian_metric}
\Vert X \Vert^2  ={\text{\small det}X}\;,
\ee
such that both ${\cal M}_{\t},{\cal M}_{\l} $ have unit spacelike length.

To make the analysis more explicit, let us assume a certain orientation of the string network, given by
$q_1 p_2 - p_1 q_2 >0$. To ensure the irreducibility of the string network made of 
the $(q_1,q_2)$ and the $(p_1,p_2)$ five-branes, we will further 
require $q_1p_2 -q_2 p_1 = 1$, namely that the corresponding $2\times 2$ matrix
$$
\G =\bem q_1&q_2 \\ p_1 & p_2\eem
$$
is an $SL(2,\Z)$ matrix \cite{Dabholkar:2006bj}.
The generalization to the charges with $\G \in GL(2,\Z)$, including 
the opposite orientation of the string network with $q_1 p_2 - p_1 q_2 =-1$, is a straightforward modification of the following discussion and will not be separately discussed here\footnote{
 It simply involves exchanging $\t$ and $\bt$ in equations (\ref{eq_kaehler_parameter})-(\ref{theta}),(\ref{mass1}),(\ref{coord_1}),(\ref{J_integral}),(\ref{eq_Omega}),(\ref{area_J_integral}).} .

\begin{figure}
\centering
\includegraphics[width=15cm]{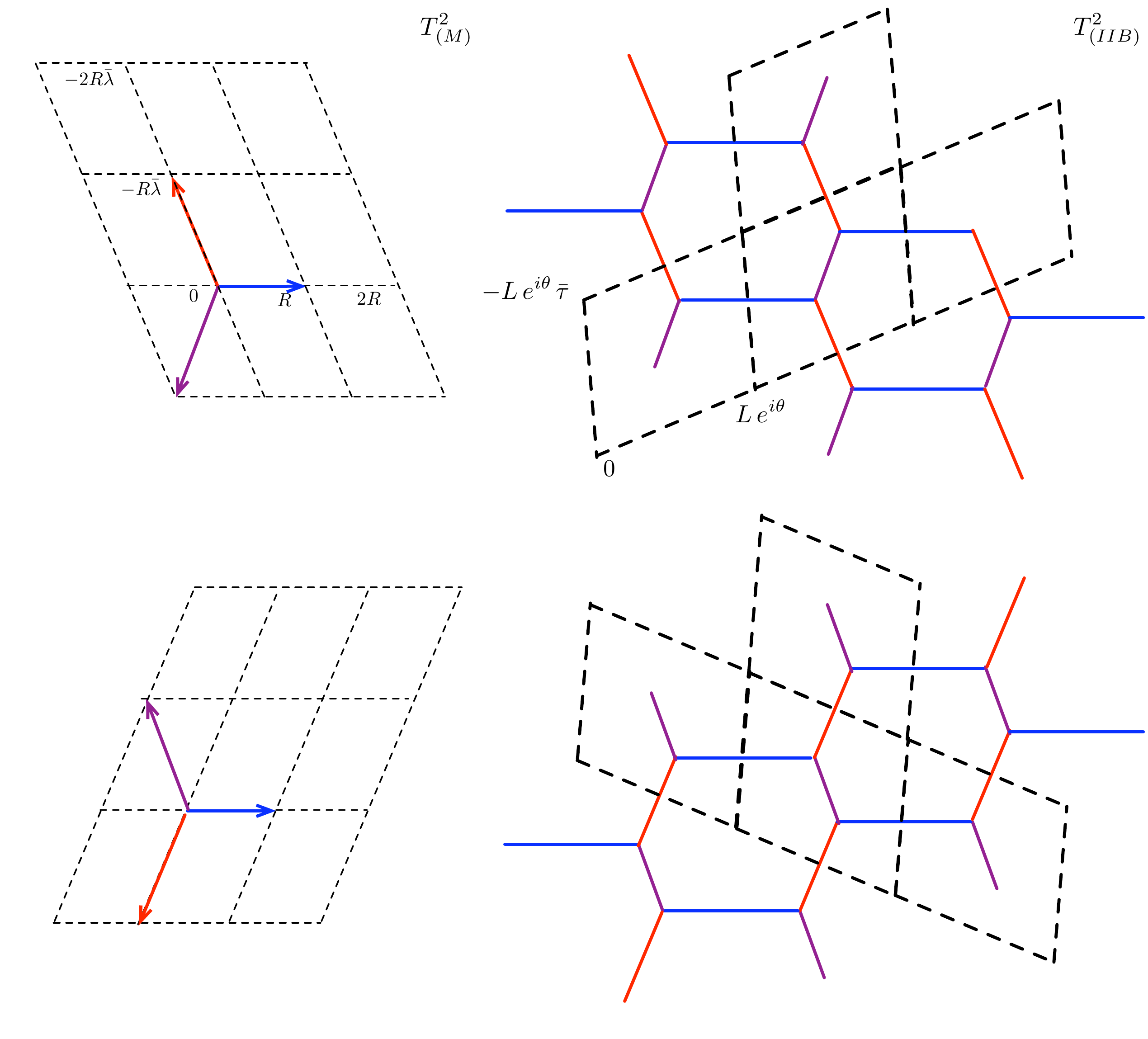} 
\setlength{\abovecaptionskip}{5pt}
\caption{\label{network_fig}\footnotesize{Two examples, described in (\ref{network_1}) and (\ref{network_2}), of the effective string network with $(q_1,q_2)=(1,0)$ and $(p_1,p_2)=(0,1)$. As discussed in  (\ref{length_general}), depending on the moduli, these networks may or may not be realized. }}
\setlength{\belowcaptionskip}{5pt}
\end{figure}

Simple kinematic consideration, or relatedly supersymmetry, requires 
that the three lines meeting at a vertex satisfy the following constraints \cite{Sen:1997xi}. The angles formed by the three legs meeting at a vertex in the periodic string network must be the same as the angles formed by the three tension vectors (\ref{central_charge_vector}) of the corresponding charges in a complex plane. 
Two examples are shown in Fig \ref{network_fig}. 

As we shall see shortly, how the supersymmetric network will be realized depends on the background moduli of the theory.  For the time being, let us focus on the one specific case depicted in the first figure in Fig \ref{network_fig}. In this case the statement about the angles simply means the following. If we view the compactification torus as $\C/R_B(\Z-\bt\Z)$ and draw the network on the same complex plane, the three vectors  
$\ell_{1,2,3}\in \C$ in this periodic network are given by 
\be\label{network_1}
\ell_1 = t_1 (T_Q+ T_P)\quad,\quad\ell_2 =t_2 T_P \quad,\quad\ell_3 = \,t_3 T_Q\;,
\ee
where the tension vectors $T_{Q,P}$ are given in (\ref{central_charge_vector}) and $t_{1,2,3}\in\R_+$ are the length parameters given by the background moduli in a way we will now describe. 

The fact that this network fits in the geometric torus $T^2_{(IIB)}$ means the length parameters satisfy
\be\label{eq_kaehler_parameter}
\bem t_1+ t_3 & t_1 \\ t_1 & t_1+ t_2\eem\bem T_Q \\ T_P \eem = 
\bem t_1+ t_3 & t_1 \\ t_1 & t_1+ t_2\eem\G \bem 1 \\ -\bar{\l}\eem = e^{i\th} R_B \bem 1 \\ -\bt\eem
\ee
for some angle $\th$ as shown in Fig \ref{network_fig}. The obvious fact that $$ (T_Q+ T_P) \bar\ell_1 + T_P \bar\ell_2+ T_Q \bar\ell_3  \in \R_+$$ then gives
\be\label{theta}
\th = \text{Arg} 
(T_Q-\t T_P)\;.
\ee
The mass of the string network, which is given by the sum of the product of the length of the legs in the type IIB torus and their respective tension, is then given by 
\be
\label{mass1}
M_{\SS IIB} =(T_Q+ T_P) \bar\ell_1+T_Q \bar\ell_2 + T_P \bar\ell_3  =  R_B V_{K3}\l_2\,\rvert \,T_Q-\t T_P\,
  \lvert\,.
\ee

Furthermore, by first solving (\ref{eq_kaehler_parameter}) for the simplest case with $\G = \mathds 1_{2\times 2}$ and considering other solutions related to it by a type IIB S-duality (\ref{IIB_S_dual}), we obtain the expression for the lengths of the three different legs in the string network
\be
\bem t_1+ t_3 & t_1 \\ t_1 & t_1+ t_2\eem = \sqrt{\frac{R_B^2\t_2}{\l_2}}\,\frac{{\cal M}_\t^{-1} + (\G^{-1})^T {\cal M}_\l \G^{-1}}{\Vert {\cal M}_\t^{-1} + (\G^{-1})^T {\cal M}_\l \G^{-1}\Vert}\;.
\ee

While the quantity on the right-hand side depends on our specific choice among charges lying on the same T-duality orbit and furthermore its derivation is only valid in the part of the moduli space with $V_{K3} \gg 1$, in what follows we shall see how this quantity can naturally be written as an T-duality invariant expression which is well-defined for general values of moduli.  

Recall that, from the four-dimensional macroscopic analysis we know the BPS mass of a dyon should be expressed in 
terms of the charges and the moduli in a specific way \cite{Cvetic:1995bj,Cheng:2008fc}. 
Especially, in the heterotic frame it depends on the right-moving 
charges only, which can be combined into the following T-duality invariant matrix
\be
\Lambda_{Q_R,P_R}  = \bem Q_R \cdot Q_R & Q_R \cdot P_R
\\Q_R \cdot P_R & P_R \cdot P_R
\eem 
\ee
and further combined with the heterotic  axion-dilaton into the 
matrix
\bea
\label{mod_vec}
{\cal Z} &=& 
 \frac{1}{\t_2} \bem 1 & -\t_1 \\ -\t_1 & 
|\t|^2\eem + \frac{1}{\Vert\Lambda_{Q_R,P_R} \Vert}\bem P_R \cdot P_R& -Q_R \cdot P_R
\\-Q_R \cdot P_R & Q_R \cdot Q_R \eem\;,
\eea
which is again invariant under T-duality transformation.

In terms of these $2\times 2$ matrices, the mass in string frame is 
given by 
\bea\notag
M^2_{\SS IIB} &=& V_{K3}\, R_B^2\, \l_2^2 \bigg( |Q_R -\bt P_R|^2 +2 
\t_2\,\Vert\L_{Q_R,P_R}\Vert\bigg)\\ \label{mass_sugrav}
&=&  V_{K3}\, R_B^2\, \t_2 \l_2^2\, \Vert \L_{Q_R,P_R}\Vert\, \Vert{\cal Z}
\Vert^2\;.
\eea

Comparing with the mass formula for the string network (\ref{mass1}), 
we can read out the expression for $Q_R,P_R$ 
\be
\Vert\Lambda_{Q_R,P_R}\Vert  = V_{K3} \l_2 = V_{K3}^{\SS (P)}\quad,\quad\frac{\Lambda_{Q_R,P_R} }{\Vert\Lambda_{Q_R,P_R}\Vert }= \G {\cal M}_\l^{-1} \G^T\;,
\ee
and thus
\be
{\cal Z} 
={\cal M}_\t^{-1} + (\G^{-1})^T {\cal M}_\l \G^{-1} \;.
\ee

From this we see that the moduli vector ${\cal Z}$ has the following two physical roles in the type IIB supersymmetric string network. First its length gives the mass of the network as in (\ref{mass_sugrav}). Furthermore its direction dictates the relation between the lengths of various legs of the network by
\be\label{length_parameters_1}
\bem t_1+ t_3 & t_1 \\ t_1 & t_1+ t_2\eem  = \sqrt{\frac{R_B^2\t_2}{\l_2}}\, \frac{{\cal Z}}{\Vert {\cal Z}\Vert}\;.
\ee

But there is clearly a problem with this formula. As the reader might have noticed, the above formula is devoid of a geometric meaning when one or more of the length parameters $t_i$ is negative. To take the simplest example, while the diagonal terms of the matrix $\cz$ are manifestly positive (\ref{mod_vec}), the off-diagonal term can be of either sign. It means that when the entries of $\cz$ fail to be all positive, for example, the network we have just described cannot exist. 

The solution to this problem is the following. As we have mentioned earlier, there are more than just one possible way to realize a supersymmetric string network with given $4D$ charges. For illustration let's now consider the following example. 
Writing 
$
\cz =\big(\begin{smallmatrix} z_{1} & z \\ z & z_2 \end{smallmatrix} \big)
$
and assume $z_1,z_2 > -z >0$ such that the network we discussed above does not exist, we will now see that the network is realized as a periodic honeycomb network with three legs given by
\be\label{network_2}
\ell_1 = t_1 (T_Q- T_P)\quad,\quad\ell_2 =-t_2 T_P \quad,\quad\ell_3 = \,t_3 T_Q
\;.
\ee
Repeating the same analysis as before we obtain the same expression for the angle $\th$ which measures the ``tilt" of  the network (\ref{theta}) and the mass of the network (\ref{mass1}), but now the length parameters are given instead by 
 \be\label{eq_kaehler_parameter_2}
\bem t_1+ t_3 & -t_1 \\-t_1 & t_1+ t_2\eem  = \sqrt{\frac{R_B^2\t_2}{\l_2}}\, \frac{{\cal Z}}{\Vert {\cal Z}\Vert}\;.
\ee
It is then easy to see that the above network (\ref{network_2}), shown in the second figure in Fig \ref{network_fig},  does exist for the range of moduli space $z_1,z_2 > -z >0$ that we consider. 

In general, as will be discussed in details in section \ref{Moduli Space as the Dual Graph}, for any arbitrary point in the moduli space, exactly one network which is given by effective strings with charges $aQ+bP$ and $cQ+dP$ wrapping the cycles $dA-cB$ and $-bA+aB$, will be realized. Here we again use $A$ and $B$ to denote the $A$- and $B$-cycle of the compactification torus $T^2_{\SS (IIB)}$. 
And the integers
$$
\g= \bem a & b \\ c&d\eem \in GL(2,\Z) 
$$
are determined by the value of moduli, which is given by the values of $\l,\t$ in the five-brane system we consider. Recall that the requirement that the inverse of an element in $GL(2,\Z)$ is again an element of the same group means that the matrix $\g$ must have determinant $\pm 1$.

In more details, the periodic network will consist of three legs given by
\be
\ell_1 =t_1\,\big(  (a+c)T_Q + (b+d)T_P \big) \quad,\quad\ell_2 =t_2\, (cT_Q+dT_P) \quad,\quad\ell_3 = \,t_3\, (a T_Q+ bT_P)
\ee
 with length parameters given by
\be\label{length_general}
\bem t_1+ t_3 & t_1 \\t_1 & t_1+ t_2\eem  = \sqrt{\frac{R_B^2\t_2}{\l_2}}\, \frac{(\g^{-1})^T{\cal Z}\g^{-1}}{\Vert {\cal Z}\Vert}\;.
\ee
As will be explained in more details in section \ref{Moduli Space as the Dual Graph}, for a given point in the moduli space, the integral matrix $\g$ has to satisfy the requirement that the above equation has a  solution with $t_{1,2,3} \in \R_+$.

\section{The Riemann Surface}
\label{The Riemann Surface}
\setcounter{equation}{0}

Following the idea of \cite{Gaiotto:2005hc} and adopting the approach of \cite{Banerjee:2008yu}, in this section we study the holomorphic embedding of a Riemann surface wrapped by the M5 brane in Euclidean M-theory which makes up the $1/4$-BPS dyons of the theory.  In particular, following \cite{Banerjee:2008yu} we write down the period matrix of such a surface for generic values of the moduli of the theory, and discuss the relationship between the degeneration of the surface and the crossing of walls of marginal stability where some dyon states might become unstable.

In order to compute the dyon partition function of the compactified type IIB theory discussed in the previous section, it is necessary to go to the Euclidean spacetime with a Euclidean time circle. 
Now recall that type IIB compactified on a circle is equivalent to M-theory 
compactified on a torus, which we will refer to as the ``M-theory torus" $T^2_{\SS(M)}$, by a T-duality transformation followed by a lift to eleven dimensions. In particular, letting the eleventh-dimension circle to have asymptotic radius $R_M$, the complex moduli and the area of the M-theory torus $T^2_{\SS(M)}$ are given by the type IIB axion-dilaton as $-\bl$ and $R_M^2 \l_2$.

In other words, in order to discuss the dyon partition function we consider M-theory compactified down to $\R^3$ on the internal manifold $K3 \times T^2_{\SS(M)}\times T^2_{(\SS{IIB})}$. 
Since the configuration we will be considering is the M5 brane wrapping the whole $K3$, 
we will now focus on the $ T^2_{\SS(M)} \times T^2_{(\SS{IIB})}$ 
factor whose moduli play the most important role in the rest of the paper. Clearly, it can 
be thought of as a space of the form $\C^2/{\mathbf{\L}}$, where the two complex planes can be taken to be the complex planes associated with the tori  $ T^2_{\SS(M)} $ and $ T^2_{(\SS{IIB})}$ respectively. Writing the coordinate 
of the two complex planes as $z_1=x_1+i y_1$ and $z_2=x_2+i y_2$, 
the lattice ${\mathbf{\L}}$ is generated by the following four vectors in 
\(\R^4\) parametrized by $(x_1,y_1,x_2,y_2)$: 
\begin{align}\notag
e_1 &= R_M\,(1,0,0,0)\quad\\ \notag e_2 &= R_M\,(-\re\bl,-\im\bl,0,0)\\ \notag
e_3 &=R_B\,(0,0,\re \,e^{i\th},\im \,e^{i\th})\\
\label{coord_1}
e_4 &=
R_B\,(0,0,-\re \,e^{i\th}\bt ,-\im \,e^{i\th}\bt )
\;.
\end{align}
For convenience we have chosen the coordinates of $\R^4$ such that the $Q$-string lies along the $x_2$-axis. See Fig \ref{network_fig}.

A priori there is no reason to require the two tori $T^2_{\SS(M)}$ and $T^2_{(\SS{IIB})}$  
be orthogonal to each other. A non-zero inner product in $\R^4$ between the vectors $\{e_1,e_2\}$ and $\{e_3,e_4\}$(\ref{coord_1})
corresponds to turning on timelike Wilson lines for the $B$- and $C$- 
two-form fields along the $A$- and $B$-cycles of of compactification torus $T^2_{(\SS{IIB})}$ in 
the original type IIB theory. But since they are absent in the Lorentzian type IIB theory we started with, in most of the following discussion we will assume that such a cross-term is absent.

\begin{figure}
\centering
\includegraphics[width=7cm]{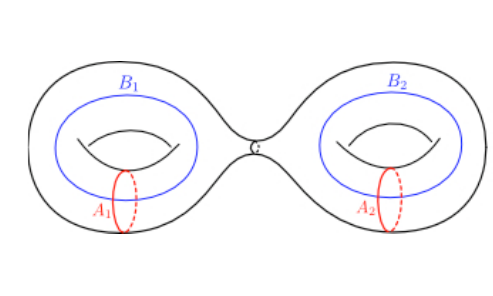} 
\setlength{\abovecaptionskip}{5pt}
\setlength{\belowcaptionskip}{5pt}
\caption{\label{genus2degenerating}\footnotesize{The degeneration of the genus two surface described in (\ref{degenerate_area}).}}
\end{figure}

After describing the M-theory set-up we now turn to 
the dyons in the theory.  The type IIB effective string network 
discussed in the previous section (\ref{charges_initial}) now becomes a genus two Riemann 
surface $\S$ inside $T^4$ upon compactifying the temporal direction and going to the M-theory frame, which has the effect of fattening the network in Fig \ref{network_fig}.  As usual, we would like to choose a canonical basis for the homology cycles of the Riemann surface $\S$ such
 that the $A$- and $B$-cycles have the following canonical
intersections:
\be\label{intersection_A_B}
A_a \cap B_b = \d_{ab}\quad,\quad A_a \cap A_b = B_a \cap B_b =0 \quad,\quad a,b=1,2\;.
\ee
We now choose the basis cycles $A_{1,2}$ and $B_{1,2}$ as shown in 
Fig \ref{genus2degenerating}. Beware that they are not directly related to the $A$- and $B$-cycles of the tori $T^2_{\SS(IIB)}$ and $T^2_{\SS(M)}$.

From the charges of the network, which translate in the geometry into the homology classes of the two-cycle  in $T^4$ wrapped by the M5 brane, we see that the Riemann surface $\S$ defines a lattice inside ${\mathbb R}^4$, with generators related to those of $\mathbf\L$ in the following way

\be\label{jacobian}
\bem \oint_{A_1} dX \\ \oint_{A_2} dX  \eem  =  \G \bem e_1 \\ e_2 \eem\quad,\quad
\bem \oint_{B_1} dX \\ \oint_{B_2} dX  \eem = \bem e_3 \\ e_4 \eem
\;.
\ee
In the above formula, $dX=(dx_1,dy_1,dx_2,dy_2)$ is the pullback on the Riemann surface $
\S$ of the one-forms on $\mathbb R^4$ in which $\S$ is embedded\footnote{For convenience and given that there's little room for confusion, here and elsewhere in this section we will not distinguish in our notation for a form in $\R^4$ and its pullback along the embedding map (\ref{embedding_map_jac}) onto the Riemann surface. }. It is easy to see that the this lattice is identical to the lattice $\mathbf\L$ (\ref{coord_1}) generated by $e_{1,\dotsi,4}$ which defines the spacetime four-torus in $\R^4$, as long as we restrict to the M5 brane charges with  $|\text{\small det}\G|= g.c.d.(Q\wedge P) =1$. We shall say more about the role of this lattice for the Riemann surface $\S$ shortly, but for that we will first need to discuss the complex structure of this surface.

The spacetime supersymmetry requires that the genus two Riemann surface to be holomorphically embedded in the spacetime $T^4$. To find the period matrix of the Riemann surface, we are interested in finding the complex structure of $\R^4$ which is compatible with the holomorphicity of $\S$. By definition this complex structure will then determine the complex structure of the Riemann surface. 
Using the natural flat metric on $\R^4$, its volume form is given by 
$$
\text{\it vol}=dx_1 \wedge dx_2 \wedge dy_1 \wedge dy_2\;, 
$$
and the space of self-dual two-forms in $\R^4$ will then be spanned by the following three two-forms
\bea\notag
f_1 &=& 
dx_1 \wedge dy_1-dx_2 \wedge dy_2\\ \notag
f_2 &=& 
 dx_1 \wedge dy_2+dx_2 \wedge dy_1\\ \notag
f_3 &=&
  dx_1 \wedge dx_2+dy_1 \wedge dy_2\;.
\eea
Recall that this three-dimensional space corresponds to the $S^2$ worth of complex structures of the hyper-K\"ahler space $\R^4$ in the following way. For a given complex structure two-form $\Upsilon$, the space of self-dual two-forms are spanned by the $(2,0)$, $(1,1)$ and $(0,2)$ form $\Upsilon=\Upsilon_1 + i\Upsilon_2 $, $J$ and $\bar{\Upsilon}=\Upsilon_1 - i\Upsilon_2 $, where $J$ is the K\"ahler form. From 
\bea
\Upsilon\wedge \bar{\Upsilon} = J\wedge J = {\text{\it vol}} \\
\Upsilon\wedge \Upsilon = \Upsilon\wedge J =0 \;,
\eea
we conclude that $J,\Upsilon_1,\Upsilon_2$ are mutually perpendicular in the pairing $\frac{\cdot \wedge \cdot }{\text{\it vol}}$ for two-forms and 
$\Upsilon_1\wedge\Upsilon_1=\Upsilon_2  \wedge\Upsilon_2 =\frac{1}{2} J\wedge J$.

If the Riemann surface $\S$ is holomorphically embedded in $\R^4$ with respect to the complex structure $\Upsilon$, the following condition is satisfied
\be
\int_\S \Upsilon = 0\;.
\ee 
To find the complex structure $\Upsilon$ compatible with the holomorphicity of $\S$ we therefore have to find a vector $J$ in the three-dimensional space of self-dual two-forms, such that the plane normal to it is the plane of all two-forms $f$ satisfying $\int_\S f =0$. This plane will then be the plane spanned by $\Upsilon_1$ and $\Upsilon_2$. From (\ref{jacobian}) we can compute the value of $f_{1,2,3}$ integrated over the surface $\S$ using the Riemann bilinear relation. From the results
\bea\notag
\int_\S {f_1} &=& 0 \\ \notag
\int_\S{f_2} &=& -R_BR_M\; \im\left( e^{-i\th} \big((q_1-\bl q_2) -\t (p_1 -\bl p_2 )\big)\right)=0\\ \notag
\int_\S{f_3} &=& R_BR_M\; \re\left( e^{-i\th} \big((q_1-\bl q_2) -\t (p_1 -\bl p_2 )\big)\right) \\ \label{J_integral}
&=& R_BR_M\;\vert (q_1-\bl q_2) -\t (p_1 -\bl p_2 )\vert\;,
\eea
 we see that the correct complex structure of $\R^4$ that gives the holomorphic embedding of the surface $\S$ is as follows 
\bea \notag
\Upsilon&=& f_1 + i f_2 = w_1 \wedge w_2 \quad,\quad w_1 = dx_1 + idx_2\;,\; w_2 = dy_1 + idy_2 \\ \label{holom_one_form}
J &=& f_3\;.
\eea
In particular, the above one-forms $w_{1},\,w_2$ form a basis of the holomorphic one-forms on the Riemann surface when pulled back along the embedding map. 
Notice that, although the above expression for the complex structure $\Upsilon$ seems to be independent of the charges and moduli, this is not quite true since we have hidden the dependence in our choice of coordinates $x_{1,2},y_{1,2}$ of $\R^4$ (\ref{coord_1}). More explicitly, one can view the complex structure as charge- and moduli-dependent through our definition of the angle $\th$
(\ref{theta}).

Now we are ready to discuss the embedding of $\S$ into the spacetime tori $ T^2_{\SS(M)}\times T^2_{(\SS{IIB})}$.
Recall that the Jacobian variety of a genus $g$ Riemann surface $\S^{\SS(g)}$ is given by the complex torus ${\cal J}(\S^{\SS(g)})=\C^g/\mathbf\L(\S^{\SS(g)})$, where $\mathbf\L(\S^{\SS(g)})$ is the lattice generated by the $2g$ vectors 
\be\label{jacobian_lattice}
\begin{array}{ccc}
&( \oint_{A_1} w_1, \dotsi, \oint_{A_1} w_g)&\\
&\vdots&\\
&( \oint_{A_g} w_1, \dotsi, \oint_{A_g} w_g)&\\
&( \oint_{B_1} w_1, \dotsi, \oint_{B_1} w_g)&\\&\vdots&\\
&( \oint_{B_g} w_1, \dotsi, \oint_{B_g} w_g)&
\end{array}
\ee
and $\{w_1,\dotsi,w_g\}$ is a basis of one-forms on the Riemann surface which are holomorphic with respect to its given complex structure. The following map, the so-called Abel-Jacobi map, then gives a holomorphic embedding of the Riemann surface $\S^{\SS (g)}$ into its Jacobian $\mathbf\L(\S^{\SS(g)})$:
\be\label{embedding_map_jac}
\varphi: \S^{\SS(g)} \to {\cal J}(\S^{\SS(g)})\quad,\quad \varphi(P)= \bem \int_{P_0}^P w_1, & \dotsi& ,\int_{P_0}^P w_g\eem,
\ee
where $P_0$ is a given arbitrary point on $\S^{\SS(g)}$. Notice that the Jacobian is defined in such a way that the above map is well-defined, namely that the images are independent of the path of integration.
In the case of our genus two surface $\S$, using the holomorphic one-forms $w_1,w_2$ given in (\ref{holom_one_form}),  from (\ref{jacobian}) we see that $\bf\L = \bf\L(\S)$, and therefore the Jacobian of the surface ${\cal J}(\S)$ is naturally identified with the spacetime $T^4$. The Abel-Jacobi map (\ref{embedding_map_jac}) therefore provides us with an explicit holomorphic embedding of the M5 brane Riemann surface $\S$ into the spacetime torus, as was suggested in \cite{Gaiotto:2005hc}.  

After discussing the complex structure and the embedding of the surface, now we are ready to compute its normalized period matrix $\O$. Consider two holomorphic one-forms 
$(\hat w_1\;\hat w_2 ) = (w_1\;w_2) V$, where $V$ is a real $2\times 2$ matrix,
such that 
\be\label{solve_V_eq1}
\bem \oint_{A_1} \hat w_1 &  \oint_{A_1} \hat w_2 \\ \oint_{A_2} \hat w_1 &   \oint_{A_2} \hat w_2 \eem =
\bem 1 &0\\ 0&1 \eem\;.
\ee
The (normalized) period matrix $\O= \re\O + i\, \im\O$ is then the symmetric $2\times 2$ matrix given by
\be
\O
=\bem \oint_{B_1} \hat w_1 &  \oint_{B_1} \hat w_2  \\ \oint_{B_2} \hat w_1& \oint_{B_2} \hat w_2   \eem =  \bem\r & \n \\ \n & \s\eem\quad
, \quad\r,\s,\n \in \C\;.
\ee

Comparing (\ref{solve_V_eq1}) and the first part of (\ref{jacobian}) one can easily obtain the explicit solution for the real matrix $V$. Integrating the resulting $\hat{w}_{1,2}$ over the B-cycles then gives $\re\O=0$, while the imaginary part of $\O$ satisfies
\be\label{eq_Omega}
\im\O\, \G\, \bem 1 \\ -\bar{\l}\eem = e^{i\th} \frac{R_B}{R_M} \bem 1 \\ -\bt\eem\;.
\ee
Up to a multiplicative factor involving the M-theory radius, this is exactly the same equation (\ref{eq_kaehler_parameter}) that the matrix of the length parameters $t_{1,2,3}$ of the type IIB string network satisfies.
We therefore conclude that the period matrix of the genus two curve wrapped by the supersymmetric M5 brane configuration is given by
\be\label{period_matrix}
\im \O =\sqrt{\frac{R_B^2\t_2}{R_M^2\l_2}}\, \frac{{\cal Z}}{\Vert {\cal Z}\Vert} \quad,\quad\re\O=0\;.
\ee

Note that the direction of the above vector in $\R^{2,1}$ is given by the 
moduli vector $\cal Z$ (\ref{mod_vec}), while the length is given  by 
the ratio of the area of the two spacetime tori. And the requirement $
\Vert\im\O\Vert\gg 1$ for rapid convergence of the partition function is 
the physical requirement that we work in the low temperature limit in the type IIB frame in which $R_B^2 \t_2 \gg 
R_M^2 \l_2$.

The fact that the period matrix is purely imaginary is really a consequence of the fact that 
our two spacetime tori $T^2_{\SS(M)}$ and 
$T^2_{\SS(IIB)}$ are orthogonal to each other, which in turn reflects the 
absence of temporal Wilson lines in the original type IIB setup. If these 
Wilson lines are turned on, the real part of the period matrix will 
instead be
\be\label{re_omega}
\re\,\O=  \bem C_{t1} & B_{t1} \\ C_{t2} & B_{t2}
\eem\G^{-1}=(\G^{-1})^{T} \bem C_{t1} & C_{t2} \\ B_{t1} & B_{t2}\eem \;,
\ee
where $B_{t1}$,$B_{t2}$,$C_{t1}$,$C_{t2}$ denote the background two-form $B$- and $C$-fields along the $A$- and $B$-cycles of the torus $T^2_{\SS(IIB)}$ and the temporal circle in type IIB. The extra condition on these Wilson lines $\mathrm{Re}\O= (\mathrm{Re}\O)^T$ could be thought of as a part of the supersymmetry condition, since if the Wilson lines do not satisfy this condition, the holomorphic embedding of the M5 brane world volume into the spacetime four-torus is not possible with respect to the given complex structure $\Upsilon$ (\ref{holom_one_form}). Put in another way, turning on the temporal Wilson lines for the two-form fields will generically change the complex structure of the surface $\S$, with exception when (\ref{re_omega}) is satisfied. But as mentioned before, in the present paper we will not consider this possibility further.
 
 Finally we would like to comment on the fact that
 the surface area of the holomorphically embedded genus two 
surface $\S$ is simply given by  
 \be\label{area_J_integral}
 A_{\SS\S} =   \int_{\S} J = \frac{i}{2}\int_{\S} (w_1\wedge \bar w_1+w_2\wedge \bar w_2) = R_M R_B\,\big\vert T_Q-\t T_P  \big\vert
 \ee
 as already computed in (\ref{J_integral}). 
 As expected, the surface area is related to the mass of the BPS object 
in the following simple way
\be
 A_{\SS\S}= \frac{R_M}{V_{K3} \l_2} M_{\SS IIB} = \frac{R_M}{V_{K3}^{\SS (M)}} 
M^{\SS (M)}
\ee
where the quantities with the superscript $\scriptstyle{(M)}$ denote the 
quantities in the M-theory unit. 

This relation between the mass and the area of the corresponding 
Riemann surface suggests a geometric way of understanding the walls of 
marginal stability, defined as the subspace in the moduli space where the BPS masses of the components of a potential bound state sum up to the BPS mass of the total charges. 
When the Riemann surface degenerates in such a 
way that it falls apart into different component surfaces which are simultaneously holomorphic, the 
area of the combined surface clearly equals to the sum of the 
area of each component surface. Upon using the above relation between the area and the BPS mass, this then directly
translates into an expected correspondence between the wall of marginal stability and wall of degeneration of the surface $\S$.
 
One simplest example of the above-mentioned phenomenon is when the genus two curve $\S$ degenerates in
such  a way that it splits from the middle and falls apart into two tori as shown in Fig 
\ref{genus2degenerating}. In this simple case, one can indeed check explicitly that the criterion 
on the period matrix for such a degeneration to happen 
is exactly the criterion that the mass, or the surface area, becomes the 
sum of the contribution of the two components
\be\label{degenerate_area}
\O = \bem \r & 0 \\ 0& \s\eem \Leftrightarrow  A_{\SS \S_1} + A_{\SS \S_2} = A_{\SS \S}
\ee
where $ A_{\SS \S_1} = \vert q_1 -\bl q_2 \vert $, $ A_{\SS \S_2} = \vert -\t(p_1 -\bl p_2) \vert $.
In other words, the above wall of marginal stability is the co-dimension one subspace of the moduli space such that the two tori defined by $\oint_{A_1} dX$, $\oint_{B_1} dX$ and $\oint_{A_2} dX$, $\oint_{B_2} dX$ respectively (\ref{jacobian}), are simultaneously holomorphic with respect to the complex structure $\Upsilon$.

To have a geometric understanding of the physics of crossing the walls 
of marginal stability, in the following section we will study the degeneration of the genus two 
Riemann surfaces of this kind in details. As we shall see, this geometric consideration will lead to 
a construction of a group of crossing the walls of marginal stability and therefore provides a geometric derivation of the group of dyon wall-crossing observed in \cite{Cheng:2008fc}.

\section{Deriving the Group of Discrete Attractor Flow}
\label{Deriving the Group of Discrete Attractor Flow}
\setcounter{equation}{0}

This section contains most important results of the present paper. In the first subsection we study a specific degeneration of  the Riemann surface and show how the effect of going through such a degeneration boils down to a change of the homology cycles. This then in turn gets translated into a change of the ``effective charges" of the system under the idenfication between the homology classes of the cycles of the surface in the internal space and the conserved charges of the system. In the second subsection we study the symmetry of the system and thereby recover the full hyperbolic reflection group underlying the structure of wall-crossing of the present theory. In the last subsection we discuss the implication of these results to the problem of enumerating supersymmetric dyonic states, and show how it leads to the prescription proposed in \cite{Cheng:2008fc} of retrieving BPS indices at different points in the moduli space from the same partition function (see also \cite{Sen:2007vb,Cheng:2007ch} for earlier discussions).

\subsection{The First Degeneration}
\label{The First Degeneration}
\begin{figure}
\centering  
\includegraphics[width=7cm]{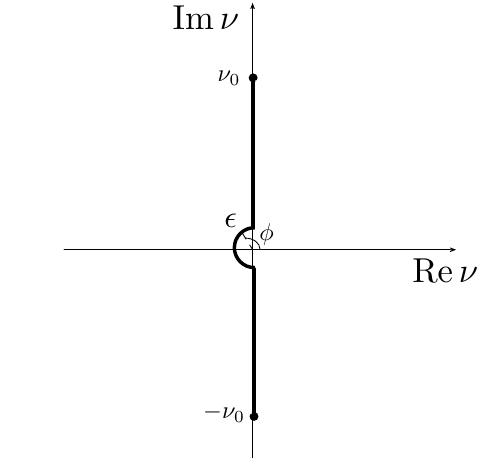}
\setlength{\abovecaptionskip}{5pt}
\setlength{\belowcaptionskip}{5pt}
\caption{\label{path}\footnotesize{In section \ref{The First Degeneration} we study the change of the Riemann surface $\S$ when its period matrix changes as (\ref{change_1})
following the above path, where $\e\to 0_+$ and $\r$ and $\s$ are held fixed at values satisfying $\im\r\,\im\s\gg (\im\n_0)^2$
.}}
 \end{figure}

First we will study what happens to the Riemann surface when the 
moduli change such that the surface goes through a degeneration mentioned at the end of the previous section. To remain in the open moduli space of the genus two Riemann surface, we study 
the change of the Riemann surface $\S$ when its period matrix $\O$ changes as 
\be \label{change_1}\bem\r & -\n \\ -\n & \s \eem\to  \bem \r & \n \\ \n & \s\eem
\ee
 following the path depicted in Fig \ref{path}. Clearly, the two end points of the path are on the different sides of the wall of marginal stability  (\ref{degenerate_area}) considered earlier. To focus on what happens to the surface when the wall is crossed, we will further zoom into the part of the path in Fig \ref{path} that is a half-circle with vanishing size:
 \be \label{half_circle}
\O=\bem\r & \e e^{i\f} \\ \e e^{i\f} &\s\eem\quad, \quad  \e\to 0_+\quad,\quad \f\in[-\frac{\p}{2},\frac{\p}{2}]\;.\ee

\begin{figure}
\centering  
\includegraphics[width=12cm]{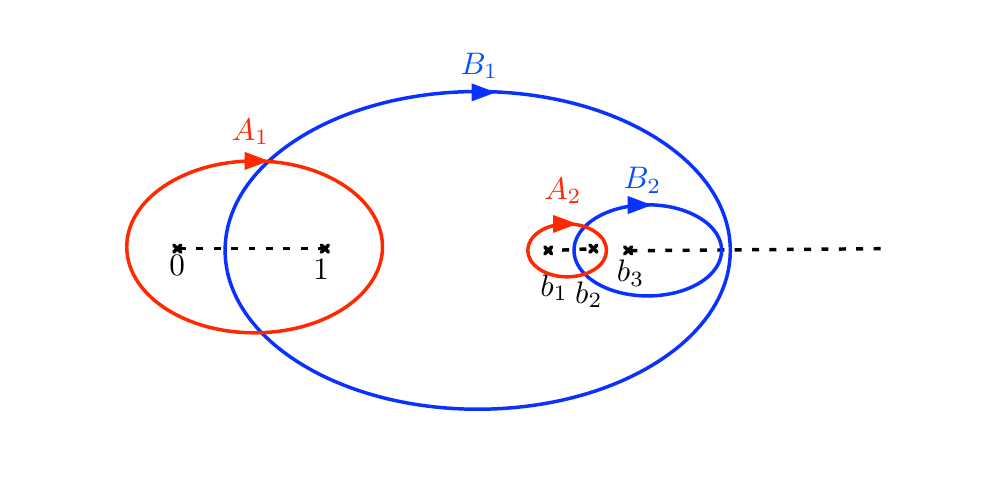}
\setlength{\abovecaptionskip}{5pt}
\setlength{\belowcaptionskip}{5pt}
\caption{\label{hyperellipticgenus2}\footnotesize{Hyperelliptic representation of the genus two surface 
  $\S$ together with a choice of its $A_i$ and $B_i$-cycles. A
  degeneration corresponding to the one shown in Fig \ref{genus2degenerating} corresponds to coalescing the branch points $b_1$, $b_2$ and $b_3$. Note that
  when we set the background two-form fields $B$ and $C$ along the timelike direction to zero, so that
  $\re\Omega=0$ (\ref{re_omega}), all branch points are colinear.}}
 \end{figure}

First recall that, every Riemann 
surface of genus two can be represented as a hyperelliptic surface with six branch points $b_{1,\dotsi,6}$
\be\label{hyperelliptic_eq}
y^2 = x (x- 1)(x- b_1)(x- b_2)(x- b_3)\;,
\ee
where we have used the conformal invariance to fix $b_{4},b_5,b_6$ to be $\inf, 0,1$ respectively.  In other words, we represent the genus two Riemann surface $\S$ as a two-sheet cover of $\C\mathbb P^1$ with six branch points $b_{1,\dotsi,6}$  and three branch cuts between $b_{2i-1}$ and $b_{2i}$ for all $i=1,2,3$, as shown in Fig \ref{hyperellipticgenus2}.

To analyze the change of the surface, in particular the homology cycles of the surface, after the imaginary part of $\nu$ changes sign, we would like to determine the normalized basis $\hat{w}_{1,2}$, satisfying (\ref{solve_V_eq1}), in terms of the local coordinate $x$ of $\C\mathbb P^1$. 

It is a familiar fact about hyperelliptic curves that the two one-forms 
$$
\frac{dx}{y},\frac{x\,dx}{y}
$$
form a basis of the holomorphic one-forms on the genus two surface $\S$ given by (\ref{hyperelliptic_eq}), see for example \cite{tata_theta}.
To achieve our goal we need to compute the integral of the above one-forms along the $A_1$, $A_2$ cycles.  First we observe that, with the choice of cycles as in Fig \ref{hyperellipticgenus2}, the integrals of a holomorphic one-form $w$ along the $A$-cycles are given by the so-called ``half-period"
\be\label{half_period}
\frac{1}{2}\oint_{A_1} w = \int_{0}^1 w \quad,\quad\frac{1}{2}\oint_{A_2} w = \int_{b_1}^{b_2} w
\ee
on the upper sheet of the hyperelliptic surface. 

To obtain an expression for these quantities in terms of the period matrix $\O$ and in particular in terms of the angle $\f$ (\ref{half_circle}), we recall that the locations of the branch points $b_{1,2,3}$ are uniquely determined by the genus two Riemann theta functions up to theta function identities \cite{tata_theta}. Explicitly, we have \cite{lebo_degeneration}
\bea
b_1 &=& \frac{\th^2[\begin{smallmatrix} 0 & 0 \\ 0&0 \end{smallmatrix}]\th^2[\begin{smallmatrix} 0 & 1 \\ 0&0 \end{smallmatrix}]}{\th^2[\begin{smallmatrix} 1 & 0 \\ 0&0 \end{smallmatrix}]\th^2[\begin{smallmatrix} 1 & 1 \\ 0&0 \end{smallmatrix}]}(0,\O)\\
b_2 &=& \frac{\th^2[\begin{smallmatrix} 0 & 1 \\ 0&0 \end{smallmatrix}]\th^2[\begin{smallmatrix} 0 & 0 \\ 0&1\end{smallmatrix}]}{\th^2[\begin{smallmatrix} 1 & 1 \\ 0&0 \end{smallmatrix}]\th^2[\begin{smallmatrix} 1 & 0 \\ 0&1\end{smallmatrix}]}(0,\O)\\
b_3 &=& \frac{\th^2[\begin{smallmatrix} 0 & 0 \\ 0&0 \end{smallmatrix}]\th^2[\begin{smallmatrix} 0 & 0 \\ 0&1 \end{smallmatrix}]}{\th^2[\begin{smallmatrix} 1 & 0 \\ 0&0 \end{smallmatrix}]\th^2[\begin{smallmatrix} 1 & 0 \\ 0&1 \end{smallmatrix}]}(0,\O)\;,
\eea
where $\th[\begin{smallmatrix} \varepsilon_1 & \varepsilon_2 \\ \varepsilon'_1 & \varepsilon'_2 \end{smallmatrix}](\zeta,\O)$ is the genus-two Riemann theta functions, defined as
$$
\th[\begin{smallmatrix} \varepsilon_1 & \varepsilon_2 \\ \varepsilon'_1 & \varepsilon'_2 \end{smallmatrix}](\zeta,\O)  = 
\sum_{n_1,n_2\in \Z}e^{2\pi i \big(\frac{1}{2}(n+\frac{1}{2} \varepsilon)^T\cdot \O\,\cdot (n+\frac{1}{2} \varepsilon)+(n+\frac{1}{2} \varepsilon)^T\cdot (\zeta+ \frac{1}{2}\varepsilon') \big)}\;,
$$
where the ``$\cdot$" denotes matrix multiplication.

While the details of these formulas are not so important for us, there are a few important immediate consequences of these expressions that we can draw. First of all, due to the fact that the genus two theta functions are a product of two genus one theta functions at leading order in $\n$ when $\n \to 0$:
$$
\th[\begin{smallmatrix} \varepsilon_1 & \varepsilon_2 \\ \varepsilon_1'&\varepsilon_2'\end{smallmatrix}] (0,\big(\begin{smallmatrix}\r&\n\\ \n & \s\end{smallmatrix}\big) )= \th[\begin{smallmatrix} \varepsilon_1 \\ \varepsilon_1' \end{smallmatrix}](0,\r)  \th[\begin{smallmatrix} \varepsilon_2 \\ \varepsilon_2' \end{smallmatrix}](0,\s) \, \big(1 + {\cal O}(\n^2)\big)\;,
$$
the three branch points coalesce when $\n\to 0$
\be
\label{coalesce}
b_1,b_2,b_3 \to b_0 = \Big(\frac{\th[\begin{smallmatrix}0\\0\end{smallmatrix}](0,\r)}{\th[\begin{smallmatrix}1\\0\end{smallmatrix}](0,\r)}\Big)^4\;.
\ee

Furthermore, from the definition of the genus two theta functions we see that 
\be
\frac{\pa}{\pa \n} b_i\big\lvert_{\n=0} = 0 \quad,\quad  i=1,2,3\;.
\ee
Therefore, for the period matrix on the half-circle given by (\ref{half_circle}) and in Fig \ref{path}, we have 
\be
b_i = b_0 + \e^2 e^{2i\f} k_i + {\cal O}(\e^4)\quad, \quad k_i = \frac{1}{2}\frac{\pa^2}{\pa \n^2} b_i\big\lvert_{\n=0} \,\in \C\;,\;\; i=1,2,3\;.
\ee
In particular, the branch points go through a $2\p$ rotation under a change $\f \to \f+ \p$. In other words, the branch points return to themselves while the period matrix undergoes a change $\n\to -\n$. 

Now we can use the above expression for the branch points near the degeneration point and (\ref{half_period}) to compute the periods along the $A_i$-cycles of the holomorphic one-forms $\frac{dx}{y}$, and $\frac{xdx}{y}$, and 
obtain the following expression for the normalized holomorphic one-forms satisfying (\ref{solve_V_eq1})
\bea\label{normalized_one_forms_hyperelliptic}
\hat w_1 &=& \frac{-1}{2(\a b_0 -\b)}\frac{(x-b_0) dx}{y} \,\big(1+ {\cal O}(\e^2) \big)\\
\hat w_2 &=&\e e^{i\f}\frac{1}{2\g(\a b_0 -\b)}\frac{(\a x-\b) dx}{y} \,\big(1+ {\cal O}(\e^2) \big)\;,
\eea
where $\a,\b,\g$ are $\f$-independent, order one constants
\bea\notag
\a &=& \int_0^1 \frac{dx}{\sqrt{x(x-1)(x-b_0)^3}} \\ \notag
\b &=& \int_0^1 \frac{xdx}{\sqrt{x(x-1)(x-b_0)^3}}\\ \notag
\g& =&\frac{1}{\sqrt{b_0(b_0-1)(k_2-k_1)}}\,\int_{0}^1\,\frac{dx}{\sqrt{x(x-1)(x-\frac{k_3-k_1}{k_2-k_1})}} \;.
\eea

While the precise values of these constants are not important for us, the above expression (\ref{normalized_one_forms_hyperelliptic}) immediately shows that, when  $\im\n$ changes sign by a $\f$ to $\f+\p$ rotation, the normalized holomorphic one-forms change like
\be
\bem \hat{w}_1\\\hat{w}_2 \eem \to \bem \hat{w}_1\\-\hat{w}_2 \eem\;
\ee
as linear combinations of the holomorphic one-forms $\frac{dx}{y}$ and $\frac{xdx}{y}$,
despite of the fact that the three coalescing branch points $b_{1,2,3}$ simply return to the original locations after a $2\p$ rotation.

This suggests that, in a representation of the hyperelliptic surface in which the holomorphic one-forms are held fixed, the homology cycles go through the following transformation 
\be\label{basis_transformation_1}
\bem A_1 \\ A_2 \eem \to \bem A_1 \\ -A_2 \eem\quad,\quad\bem B_1 \\ B_2 \eem \to \bem B_1 \\ -B_2 \eem\;.
 \ee
Indeed, it is not difficult to check  that the periods of any holomorphic one-form $w$ along $A_2$ and $B_2$ cycles
$$
\frac{1}{2}\oint_{A_2}w=  \int_{b_1}^{b_2}w\;\;,\;\;\frac{1}{2}\oint_{B_2}w=  \int_{b_2}^{b_3}w
$$
change sign under $\f\to \f+\p$. 

Another way to understand this change of homology basis is the following. From the expression of the normalized holomorphic one-forms (\ref{normalized_one_forms_hyperelliptic}) we see that, to the leading order in $\e$ we have the two separated genus one surfaces described by
\be\label{two_tori_split}
y'^2 = x(x-1)(x-b_0) \;\;,\;\; y''^2= (x-b_1)(x-b_2)(x-b_3)\;.
\ee
Indeed, from the following relationship between the cross-ratio of the four branch points ${\mathfrak b}_{1,2,3,4}$ of a genus one surface and the torus  complex moduli $\til\t$ \cite{tata_theta}
\be
\frac{({\mathfrak b}_3-{\mathfrak b}_1)({\mathfrak b}_4-{\mathfrak b}_2)}{({\mathfrak b}_2-{\mathfrak b}_1)({\mathfrak b}_4-{\mathfrak b}_3)}= \Big(\frac{\th[\begin{smallmatrix}0\\0\end{smallmatrix}](0,\til\t)}{\th[\begin{smallmatrix}1\\0\end{smallmatrix}](0,\til\t)}\Big)^4
\ee
one can check that the following two genus-one curves have complex moduli equal to $\r$ and $\s$ respectively.
From the above expression (\ref{two_tori_split}) it is manifest that, when $b_i$'s go through a $2\p$ rotation around their common converging point $b_0$, nothing happens to the first genus one surface while the second one goes through a sheet exchange (or ``hyperelliptic involution") $y'' \to -y''$ corresponding to the monodromy
\be
\bem A_2 \\ B_2 \eem \to \bem -A_2 \\ -B_2 \eem\;.
\ee
The latter can be explicitly seen by substituting
\be
x = b_0 + e^{2i\f} \til x \;\;,\;\; y'' = e^{3i\f} \til y\;
\ee
in the second equation of (\ref{two_tori_split}).

In general, when we change the basis such that the $A_i$-cycles are changed to 
\be\label{change_basis_A}
\bem A_1 \\ A_2 \eem \to \bem a & b \\ c& d \eem \bem A_1 \\ A_2 \eem\quad,\quad \g = \bem a & b \\ c& d \eem \in GL(2,\Z)\;,
\ee
the corresponding change of the $B_i$-cycles is then fixed by the canonical intersection (\ref{intersection_A_B}) to be  
\be\bem B_1 \\ B_2 \eem \to \pm \bem d& -c\\ -b & a\eem\bem B_1 \\ B_2 \eem= (\g^T)^{-1} \bem B_1 \\ B_2 \eem\;,
\ee
where the $\pm$ signs are taken when $ad-bc=\pm1$. 
Under this transformation, the period matrix transforms as  
\be\label{shorthand}
\O \to \g(\O)\equiv(\g^{-1})^T \O \g^{-1}.\ee

Without changing the Riemann surface, such a change of basis has an interpretation as performing a physical S-duality in the heterotic frame, extended with the $\Z_2$ spacetime parity exchange. To see this, first inspect the expression (\ref{jacobian}) for the vectors defining the Jacobian of the surface. The effect of the above change of basis 
on these vectors is equivalent to the following heterotic S-duality transformation, or equivalently the modular transformation of the torus in the type IIB frame
\be\label{S_dual}
\bem Q \\ P \eem \to \bem a & b \\ c & d \eem  \bem Q \\ P \eem\quad,\quad \t \to \frac{a\t + b}{c\t+d} \quad \text{  or    } \quad \frac{a\bar\t + b}{c\bar\t+d} \quad
\text{   for   } \quad ad-bc = \pm 1\;\;
\ee
with the corresponding change of $R_B$ such that the area of the type IIB torus remains invariant. In particular, the fact that the moduli vector ${\cal Z}$ transforms as ${\cal Z} \to (\g^{-1})^T {\cal Z} \g^{-1}$ under the above S-duality transformation is then consistent with the transformation of the period matrix under a change of homology basis.

Now let's go back to the evolution (\ref{half_circle}) of the Riemann surface through the degeneration wall $\n=0$, due to the corresponding change of the moduli vector ${\cal Z}$ (\ref{period_matrix}). What we have seen can be summarized as follows: when the moduli change across the wall of marginal stability following the path corresponding to an angle-$\p$ rotation of the phase of $\n$ (\ref{half_circle}), the holomorphic one-forms of the surface $\S$ change in such a way that all their $A_2$, $B_2$ periods change signs while their periods along the $A_1$, $B_1$ cycles remain the same. This is equivalent to keeping the surface unchanged but change the basis for the homology cycles in the way (\ref{change_basis_A}) 
given by the following element in $GL(2,\Z)$
\be\label{def_R}
R = \bem  1 & 0 \\ 0&-1 \eem\;.
\ee
In other words, the process of keeping the charge fixed while varying the moduli across the wall of marginal stability following (\ref{half_circle}) is equivalent to keeping the moduli vector $\cz$ unchanged but changing the charges 
\be\label{new_charge_1}
\bem Q \\ P \eem \to R \bem Q \\ P \eem\;. 
\ee

 This observation has the following implication for the counting of the BPS states. Consider the partition function of the theory
 $$ Z(\O) = \sum_{P^2,Q^2,P\cdot Q} (-1)^{P\cdot Q +1} D(P,Q) e^{i\p (\r P^2 + \s Q^2 + 2\n P\cdot Q)} $$
 which is a path integral computed on the Riemann surface $\S$, it clearly depends on its period matrix $\O$ and therefore on the moduli vector ${\cal Z}$ through its relation to the period matrix (\ref{period_matrix}). When the moduli change in such a way that the Riemann surface goes through a degeneration described in  (\ref{degenerate_area}), from the above reasoning we see that the partition function remains unchanged while a transformation of ``effective charges" given in (\ref{new_charge_1}) has to be performed. This corresponds to the change of the highest weight of the Verma module as described in \cite{Cheng:2008fc}. 
  
 It is also easy to understand the nature of this degeneration in the  type IIB five-brane network picture.
From the relation between the period matrix of the genus two curve in M-theory and the length parameters for the periodic string network (\ref{length_general},\ref{period_matrix}), we see that the degeneration of the Riemann surface characterized by $\n=0$ corresponds to the degeneration of the string network characterized by $t_1=0$. For example, starting from a region in the moduli space with ${\cal Z} = \big(\begin{smallmatrix}z_1 & z\\ z&z_2\end{smallmatrix}\big)$ with $z_1,z_2>z>0$,
what happens when $t_1=0$ is a transition from the network described by (\ref{network_1}), or the first figure in Fig \ref{network_fig}, to the network described by (\ref{network_2}), or the second figure in Fig \ref{network_fig}. The above claim that the final surface has the same period matrix under a change of homology basis corresponding to (\ref{new_charge_1}), is then reflected by the fact that the two defining equations (\ref{network_1}),(\ref{network_2}) transform into each other under the transformation of the charges (\ref{new_charge_1}).

\subsection{The Symmetry of the Weyl Chamber}
 \label{The Symmetry}
 
 In the previous subsection we have studied in detail a particular degeneration of the Riemann surface and what it implies for the index counting the BPS states under the crossing of the corresponding wall of marginal stability. In this subsection we will turn to studying the symmetry of the system and see how it will help us to uncover the full structure of the group of wall-crossing of the theory. 
 
 First we note that, under our convention that the two $A$-cycles of the surface are chosen to circle two of the three pairs of branch points $\{b_{2i-1},b_{2i}\}$, the choice shown in Fig \ref{hyperellipticgenus2} is not quite unique. In other words, from all the possible change of basis of the form (\ref{change_basis_A}), the exchange and permutation of the cycles $A_1,A_2, -A_1-A_2$ correspond to a symmetry of our hyperelliptic model (\ref{hyperelliptic_eq}) in that we do not need to change the set of branch points $\{b_1,\dotsi,b_6\}$ in order for the new $A_i$-cycles to again circle the cuts joining the pairs  $\{b'_{2i-1},b'_{2i}\}$ of the new branch points. 
We therefore conclude that there is a symmetry group with six elements acting on the hyperelliptic surface (Fig \ref{hyperellipticgenus2}), corresponding to six ways of associating the three cuts joining $\{b_{2i-1},b_{2i}\}, i=1,2,3$,  to the three homology cycles $(A_1,A_2, -A_1-A_2)$. From the above discussion we see that this group $D_3\subset GL(2,\Z)$ is the same as the symmetry group of a regular triangle, generated by the order two element which acts on the period matrix as
\be\label{action_D3_1}
\O \to RS(\O)
\ee
and which corresponds to $(A_1,A_2, -A_1-A_2)\to(A_2,A_1, -A_1-A_2)$, together with the order-three element which acts as
 \be\label{action_D3_2}
\O \to ST(\O)
\ee
and corresponds to $(A_1,A_2, -A_1-A_2)\to(A_2,-A_1-A_2,A_1 )$, where $T$ and $S$ denotes the usual T- and S- transformation matrix $\big(\begin{smallmatrix}1&1\\0&1\end{smallmatrix}\big)$ and $\big(\begin{smallmatrix}0&1\\-1&0\end{smallmatrix}\big)$, while $R$ was already given in (\ref{def_R}). 
Also here we have used the shorthand notation introduced in (\ref{shorthand}). 

\begin{figure}
\centering
\includegraphics[width=15cm]{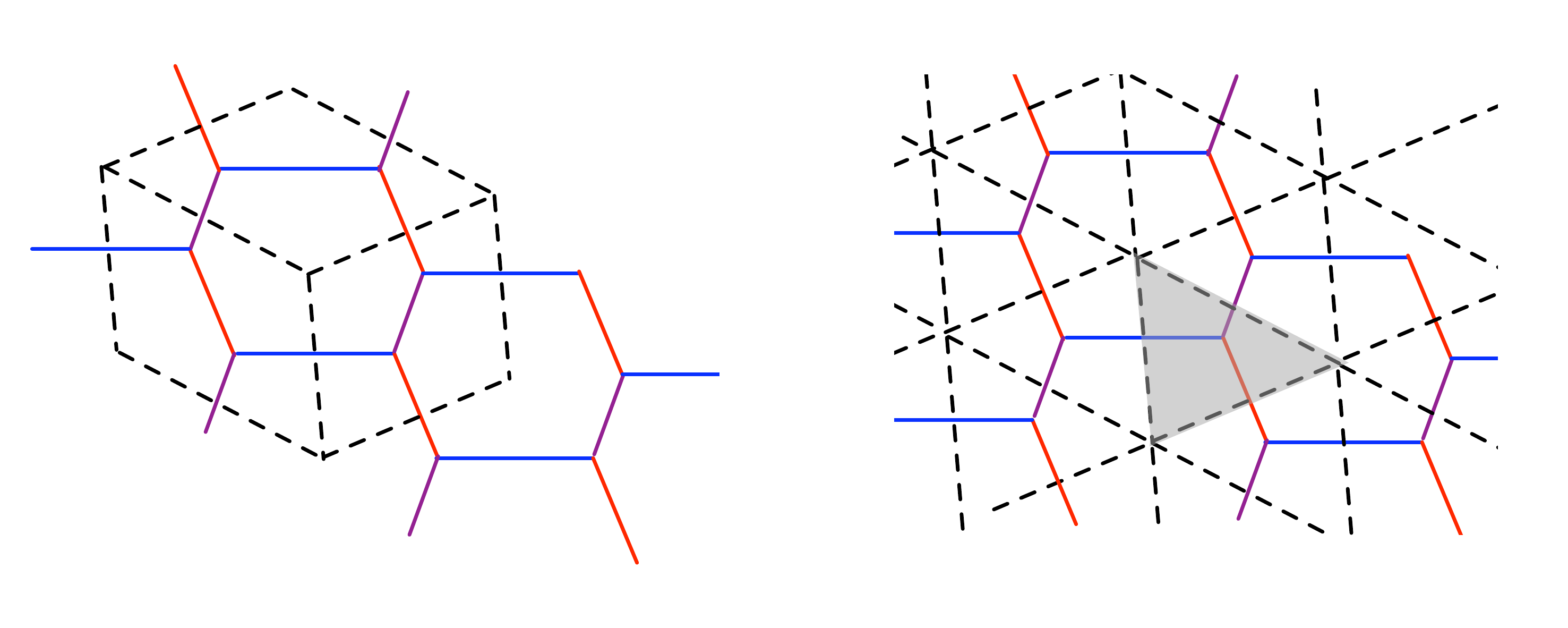} 
\setlength{\abovecaptionskip}{5pt}
\caption{\label{dual_graph}\footnotesize{{\bf(i)} Different possible ways of compactifying the periodic network on a torus. {\bf(ii)} The moduli space as the dual graph of the five-brane network. }}
\setlength{\belowcaptionskip}{5pt}
\end{figure}

The existence of this symmetry is even more apparent in the type IIB picture of five-brane network. For concreteness of the discussion we will now assume that $\cz = \big(\begin{smallmatrix}z_1&z\\z&z_2\end{smallmatrix}\big)$ satisfies $z_1,z_2 > z >0$, so that the network  shown in the first figure in Fig \ref{network_fig} is realized. But, suppose that we are given this periodic network given by (\ref{network_1}), there is actually more than one way to fit it into a parallelogram tessellation of the plane. In other words,  there is in fact more than one torus compactification of the network possible. From the fact that the vertices of the parallelogram lie at the center of the honeycomb lattice, we conclude that there are three such parallelogram tessellations possible, as shown in Fig \ref{dual_graph} as resembling the three sides of a three-dimensional cube. These three parallelograms then give six possible tori (for each parallelogram we have two ways of choosing the $A$- and $B$-cycle), corresponding to $3!=6$ ways of placing the charge labels $(Q,P,-Q-P)$ to the three legs of the network. This singles out a six-element subgroup of the extended type IIB modular group (or the heterotic S-duality group) $GL(2,\Z)$ (\ref{S_dual}). Not surprisingly, this is exactly the same $D_3$ we discovered earlier as the symmetry group of the hyperelliptic representation of the Riemann surface $\S$.  This correspondence is to be expected from the identification between the change of basis of the homology cycles on $\S$ and the heterotic S-duality discussed in the previous subsection (\ref{change_basis_A}-\ref{S_dual}).

More explicitly, from (\ref{action_D3_1},\ref{action_D3_2}) and the relation between the period matrix and the length parameters of the network  (\ref{length_parameters_1},\ref{period_matrix}) we see that the length parameters indeed transform as
\be
RS: (t_1,t_2,t_3)\to(t_1,t_3,t_2)\quad,\quad ST:(t_1,t_2,t_3)\to(t_2,t_3,t_1)
\ee
under the action of the order two and three generators of the symmetry group $D_3$. 

As mentioned earlier, this symmetry group is the symmetry group of a equilateral triangle. Geometrically, the relevant equilateral triangle here cannot be literally the triangle in the dual graph of the five-brane network as shown in Fig \ref{dual_graph}, since in general there is no reason to expect them to be equilateral using the flat metric on the plane. This inspires us to take a closer look into the matrix of length parameters, which can be written in a way which makes the  $D_3$ symmetry manifest
\be
R_M\,\im \O=\bem t_1+ t_3 & t_1 \\ t_1 & t_1+ t_2\eem= \frac{t_2+t_3}{2}\a_1+\frac{t_1+t_3}{2}\a_2+\frac{t_1+t_2}{2}\a_3
\ee
where 
\be\label{alphas}
\a_1 = \bem0&-1\\-1&0\eem\;,\;\a_2 = \bem2&1\\1&0\eem\;,\;\a_3 = \bem0&1\\1&2\eem\;.
\ee
This natural basis $\{\a_{1,2,3}\}$ has the following matrix of inner products using the standard $GL(2,\Z)$-invariant Lorentzian metric (\ref{Lorenztian_metric})
\be\label{cartan}
-2\, (\a_i,\a_j) = \bem 2 &-2 &-2\\-2 &2 &-2\\-2 &-2 &2 \eem
\ee    
and therefore forms a equilateral triangle in the hyperbolic space $\R^{2,1}$. 
The group $D_3$ which permutes $\a_{1,2,3}$ can therefore be thought of the symmetry group of this equilateral triangle. We note that the basis $\{\a_{1,2,3}\}$ we used above is exactly the basis for the roots of the Borcherds-Kac-Moody algebra adopted in \cite{Cheng:2008fc}, and in particular the matrix of inner products in (\ref{cartan}) is simply the real part of the Cartan matrix of the algebra.

To summarize, we have found in this subsection a six-element symmetry group $D_3$ of the dyon system at a given point in the moduli space which is evident in both the M-theory Riemann surface picture as well as the type IIB network picture. In the following subsection we will use this symmetry to find all the generators of the hyperbolic reflection group $W$ playing the role of the group of wall-crossing in the $\cn=4$ theory we discussed, and subsequently derive the full group structure of the dyon BPS index.

\subsection{Moduli Space as the Dual Graph}
\label{Moduli Space as the Dual Graph}

In subsection \ref{The First Degeneration} we have studied in details the change of the Riemann surface across a degeneration point (\ref{half_circle}) where the surface falls into two separate tori as depicted in Fig \ref{genus2degenerating}. 
From the above discussion about the symmetry of the system, we see that there are two more natural degenerations of the genus two Riemann surface $\S$ we should consider. The corresponding transformation of the period matrix is simply the transformation (\ref{change_1}) of the first degeneration we have studied in subsection \ref{The First Degeneration}, now conjugated with elements of the symmetry group $D_3$. 
From (\ref{action_D3_1},\ref{action_D3_2}), we see that apart from the group generator $w_1=R$  (\ref{new_charge_1}), we should also consider the generators $$w_2=(ST)^{-2}R\, (ST)^2
 \quad \text{and}  \quad w_3 = (ST)^{-1}R\,(ST)
 \;. $$ Together they generate a non-compact reflection group, which we will denote by $W$. Furthermore, it is not difficult to show  \cite{Feingold-Frenekel,Cheng:2008fc} that the extended S-duality group $PGL(2,\Z)$\footnote{Recall that $PGL(2,\Z)$ is obtained from $GL(2,\Z)$ by identifying the elements $\g$ and $-\mathds{1}\cdot \g \in GL(2,\Z)$. In the heterotic frame this corresponds to identifying two systems with the same value of axion-dilaton $\t$ and charges which are related to each other by a charge conjugation $\big( \begin{smallmatrix}Q\\P \end{smallmatrix}\big) \to \big( \begin{smallmatrix}-Q\\-P\end{smallmatrix}\big) $. In the type IIB frame this corresponds to a trivial change of basis for the homology cycles $\big( \begin{smallmatrix}A\\B \end{smallmatrix}\big) \to \big( \begin{smallmatrix}-A\\-B\end{smallmatrix}\big) $ of the compactification torus $T^2_{\SS (IIB)}$.} is a semi-direct product
\be
PGL(2,\Z) = W\rtimes D_3\;
\ee
of the reflection group $W$ and the symmetry group $D_3$.

It is clear what these three degenerations correspond to in the type IIB and as well as in the  M-theory picture. In the former case they are the three ways in which the five-brane network can disintegrate, namely letting one of the three legs having vanishing length. In the latter case, on the other hand, they correspond to coalescing the branch points $b_{3,4,5}$, $b_{5,6,1}$ or   $b_{1,2,3}$. Remember that coalescing three out of the total of six branch points is equivalent to coalescing the complementary set of three branch points.

More explicitly, these three generators $w_{1,2,3}$ of the group $W$ correspond to the following change of the network
\be
w_i : t_i\to -t_i\;,\;\quad t_i+t_j \;\;\text{invariant  for   }j\neq i\;.
\ee 
Equivalently, they can also be represented as the following reflections in the $(2+1)$-dimensional Minkowski space in which the period matrix $\im\O$ takes its value
\be
w_i :\O \to\O - 2\frac{(\a_i,\O)}{(\a_i,\a_i)}\a_i\;,
\ee 
where the basis vectors $\a_i$'s are defined in (\ref{alphas}).
Recall that we have chosen the M-theory background such that the period matrix $\O$ is purely imaginary and therefore directly related to the length parameters of the five-brane network.

Applying the same analysis as in subsection \ref{The First Degeneration} to the other two degenerations corresponding to $w_2$ and $w_3$, one can conclude that going through such a degeneration wall has the following effect on the counting of BPS states. Upon applying a suitable change of basis analogous to (\ref{basis_transformation_1}), after the degeneration we regain the original partition function but now with a different effective charges related to the original charges by 
\be\label{group_charges}
\bem Q \\ P \eem\to w_i \bem Q \\ P \eem\;.
\ee

From the above consideration, we arrive at a picture of the moduli space with its partitioning by the walls of marginal stability 
given by the the dual graph of the honeycomb lattice representing the five-brane network. This is shown in Fig \ref{dual_graph}. 
To understand this better, let's start in one of the dual triangles, let's say the gray triangle which denotes the part of the moduli space with $\cz = \big(\begin{smallmatrix}z_1&z\\z&z_2\end{smallmatrix}\big)$, $z_1,z_2 > z >0$, such that the network (\ref{network_1}) as shown in the first figure in Fig \ref{network_fig} is realized.
We shall choose it to be our ``fundamental domain" ${\cal W}$, a name that will be justified shortly.
 
A degeneration happens when one of the length parameters $t_i$'s goes to zero. As we discussed at the end of section \ref{The Riemann Surface}, this corresponds to crossing a physical wall of marginal stability. When this happens we move to the neighboring triangle, divided from the fundamental triangle ${\cal W}$ by the side of  the triangle intersecting the leg of the network whose length parameter has just goes through a zero. In this new triangle, the effective charges are related to the original one by the corresponding group element (\ref{group_charges}). For instance, associated to the triangle that shares one side with ${\cal W}$ which intersects the leg whose length parameter is denoted by $t_1$ are the effective charges $(Q,-P)$ and the region in the moduli space with $\cz = \big(\begin{smallmatrix}z_1&z\\z&z_2\end{smallmatrix}\big)$, $z_1,z_2 > -z >0$.

Given the original charges, this procedure can then be iterated. We thus conclude that each triangle of the dual graph has the effective charges $(Q_v,P_v)$ associated to it, where $v$ labels the vertices in the hexagonal lattice, or equivalently the triangles (the faces) of the dual graph, which represent the corresponding regions in the moduli space. Furthermore, in this way each of the triangles can be identified with the fundamental domain of the group $W$, generated by the three elements $w_{1,2,3}$ (\ref{group_charges}), which by construction plays the role of crossing the walls of marginal stability of the theory. 

Now that each triangle has a set of charges $(Q_v,P_v)$ associated to it, while the period matrix of the genus two Riemann surface $\S$ and therefore the partition function remains the same for each triangle, generically we conclude that there is also a different BPS index $D(Q,P)\vert_v= D(Q_v,P_v)$ associated with each triangle. The difference between $D(Q_v,P_v)$ with different $v$ has been calculated in \cite{Sen:2007vb,Cheng:2008fc} for the present theory and was shown to be consistent with the macroscopic wall-crossing formula. 

To sum up, we have derived the following one-to-one correspondence
\bea\notag
&&\text{vertex  }v  \text{  of string network } \leftrightarrow \text{a triangle in dual graph} \\ \notag
&&\leftrightarrow \text{effective charges } (Q_v,P_v) \leftrightarrow \text{BPS index  }D_v =D(Q_v,P_v)   \\ \label{conclusion}
&&\leftrightarrow \text{an element  }w_v\in W  \leftrightarrow\text{a region in the moduli space  } {\cal Z} \in w_v({\cal W})\;.
\eea
The property of the BPS dyon index of the present $\cn=4$ theory that the indices in different parts of the moduli space are given by the same partition function and have the form $D(Q,P)\vert_v= D(Q_v,P_v)$ was observed in \cite{Sen:2007vb,Cheng:2007ch} based on the macroscopic prediction for the change of index upon crossing a wall of marginal stability. And the fact that these different regions of the moduli space with different BPS indices are in one-to-one correspondence with elements of a hyperbolic reflection group $W$ is later observed in \cite{Cheng:2008fc} based on a four-dimensional macroscopic analysis. What we have seen in this paper is how these properties can be understood as the consequence of the simple consideration of the supersymmetry of the effective string network, or equivalently the holomorphicity of the M5 brane world-volume, when the limit of decoupled 4D gravity is taken.

\section{Discussion}
\label{Conclusion}

In this paper we work in the decompactification limit $(V_{K3}\!\gg\! 1)$ and 
show how the group structure underlying the moduli dependence of the dyon BPS index of the $\cn=4$ $K3 \times T^2$ compactification of type II theory can be understood as simply a consequence of the supersymmetry of the dyonic states. From the other point of view, this group structure is simply the consequence of the fact that the BPS spectrum of the theory is given by the appropriate representation of a Borcherds-Kac-Moody algebra. The Weyl group of the algebra, which is a symmetry group of the root system of the algebra, then plays the role of the group of wall-crossing for the physical degeneracy of the dyonic states \cite{Cheng:2008fc}. Therefore, we hope that the microscopic derivation of the Weyl group presented in the present paper will be the first step towards an understanding of the microscopic origin of the Borcherds-Kac-Moody algebra in the dyon spectrum. 

For this purpose, it is important to be clear about what we do not derive from the simple analysis of the present paper. First of all, while we assume that the partition function is a functional integral on the genus two Riemann surface $\S$ and therefore depends only on the period matrix of the surface, justified by the fact that an M5 brane wrapping the surface and the K3 manifold in the Euclidean spacetime is equivalent to a fundamental heterotic string whose world volume is the genus two surface \cite{Witten:1995ex}, we have not derived the partition function itself from our simple consideration. A discussion about the subtleties of computing the partition function in a very similar context can be found in  \cite{Banerjee:2008yu} and will therefore not be repeated here. Relatedly, the presence of the Borcherds-Kac-Moody algebra \cite{Dijkgraaf:1996it,Cheng:2008fc} is far from evident from our simple analysis. 

Furthermore, we have not commented on the role of the group $W$ as the group of a discretized version of attractor flows. 
As discussed in details in \cite{Cheng:2008fc}, this interpretation naturally arises due to  the existence of a natural ordering among the elements of the hyperbolic reflection group $W$, and the fact that for given total charges, there is a unique endpoint of this ordering, corresponding to the attractor point of these charges. From the point of view of the Borcherds-Kac-Moody algebra, the Verma module relevant for the BPS index is the smallest one when the moduli are at their attractor value. By working in the limit that the type IIB five-branes are all much heavier than all the $(p,q)$ strings, we have no way of telling which of the triangles in the dual graph in Fig \ref{dual_graph} contains the attractor point. This is of course consistent with the fact that our analysis in the text is independent of the values of the T-duality invariants $Q^2, P^2, Q\cdot P$ (\ref{charges_initial}), due to the decompactification limit we are taking. 
But this can easily be cured by going to the next leading order in ${\cal O}(V_{K3}^{-1})$. See also \cite{Banerjee:2008yu}. By minimizing the surface area of the genus two surface (\ref{area_J_integral}) with the volume of the two tori $R_B^2\t_2,R_M^2\l_2$ held fixed, with now the next leading corrections included, one indeed obtains the attractor equation 
$$
{\cal M}_\t = \frac{\L_{Q_R,P_R}}{\Vert \L_{Q_R,P_R} \Vert} = \frac{1}{\sqrt{Q^2 P^2 -(Q\cdot P)^2}} \bem Q\cdot Q & Q\cdot P \\ Q\cdot P & P\cdot P\eem\;,
$$
as expected. This extra piece of information will then single out a triangle in the dual graph as the attractor region and completes the interpretation of the hyperbolic reflection group derived in the present paper as the group underlying the macroscopic attractor flow of the theory.

Finally we would like to comment on the cases of other $\cn=4$ string theories.  In the present paper we have focussed on the $\cn =4$ theory of $K3\times T^2$ compactified type II theory, while from the analysis in \cite{Cheng:2008kt} we expect very similar group structures to be present also in the $\Z_n$-orbifolded theories, the so-called CHL models \cite{Chaudhuri:1995bf}, for $n<4$. For the $\Z_2$-orbifold theory, considering the double cover of the genus two Riemann surface relevant for the computation of the partition function \cite{Dabholkar:2006bj,Dijkgraaf:1987vp} and the corresponding string network, a similar analysis can be employed to understand the group structure in that case. The situation of other orbifolded theories is much less clear. 
In particular, in \cite{Cheng:2008kt} it was discovered that a similar group structure and an underlying Borcherds-Kac-Moody algebra cease to exist when $n> 4$. In particular, from the macoscopic analysis it was shown that the symmetry group of a fundamental region bounded by walls of marginal stability has infinitely many elements when $n>4$. From the analysis of the present paper, this symmetry group is expected to be the symmetry group of the dyonic string network/Riemann surface. One might thus suspect that the corresponding dyon network does not exist in the $\Z_{n>4}$ theories.
 It would be very interesting to understand the group structure of the dyon degeneracies of other orbifolded $\cn=4$ theories.

\section*{Acknowledgments}
It is a pleasure to thank Atish Dabholkar, Frederik Denef, Andy Neitzke, Alessandro Tomasiello, Erik Verlinde for very useful conversations. In particular we would like to thank Ashoke Sen for crucial discussions at the initial stage of the project. M.C. would also like to thank the hospitality of the ``Monsoon workshop on String Theory"  at TIFR and LPTHE, Jussieu, where part of the work was done. The research of M.C. is supported by the Netherlands Organisation for Scientific Research (NWO), and the research of L.H. is supported by an NWO Spinoza grant.

\end{document}